\documentclass[aps,preprint,pre,groupedaddress, showpacs]{revtex4-1}
\usepackage{amsmath,enumerate,amsfonts,color,amssymb, amsthm}
\usepackage{graphicx}
\usepackage{dcolumn}
\usepackage{bm}

\usepackage{tikz}
\usepackage{natbib}

\begin{document}

\title{Phase Transition Dynamics and Stochastic Resonance in Topologically Confined Nematic Liquid Crystals}
\author{Yucheng Hu}
\email[]{huyc@tsinghua.edu.cn}
\author{Liu Hong}
\affiliation{Zhou Pei-yuan Center for Applied Mathematics, Tsinghua University, Beijing 100084, China.}

\author{Weihua Deng}
\affiliation{School of Mathematics and Statistics, Gansu Key Laboratory of Applied Mathematics
and Complex Systems, Lanzhou University, Lanzhou 730000, China}

\date{\today}

\begin{abstract}
Topological defects resulted from boundary constraints in confined liquid crystals have attracted extensive research interests. In this paper, we use numerical simulation to study the phase transition dynamics in the context of stochastic resonance in a bistable liquid crystal device containing defects. This device is made of nematic liquid crystals confined in a shallow square well, and is described by the planar Lebwohl-Lasher model. The stochastic phase transition processes of the system in the presence of a weak oscillating potential is simulated using an over-damped Langevin dynamics. Our simulation results reveal that, depending on system size, the phase transition may follow two distinct pathways: in small systems the pre-existing defect structures at the corners hold until the last stage and there is no newly formed defect point in the bulk during the phase transition; In large systems new defect points appear spontaneously in the bulk and eventually merge with the pre-existing defects at the corners. For both transition pathways stochastic resonance can be observed, but show dramatic difference in their responses to the boundary anchoring strength. In small systems we observe a ``sticky-boundary'' effect for a certain range of anchoring strength in which the phase transition gets stuck and stochastic resonance becomes de-activated. Our work demonstrates the dynamical interplay among defects, noises, and boundary conditions in confined liquid crystals.
\end{abstract}


\maketitle

\section{Introduction}
Topological defects in liquid crystals have attracted tremendous attention in recent years. On the application side, liquid crystal devices containing defects find many important applications in fields like display~\cite{ge2009electro-optics}, laser technology~\cite{moreira2004cholesteric}, and self-assembling materials~\cite{nych2013assembly, hijnen2010self-organization}. On the theoretical side, defect in liquid crystals is an important research topic that is closely related to knot theory~\cite{martinez2014mutually}, partial differential equations~\cite{lin1999mapping}, phase transition~\cite{ondriscrawford1993curvature-induced, tomar2012morphological}, etc. Although the static properties of defects in equilibrium state have been extensively studied~\cite{ schopohl1987defect, mkaddem2000fine, hu2016disclination}, relatively little work has focused on the dynamical properties of defects and their roles in phase transition.

Stochastic resonance is a peculiar dynamical phenomenon that involves phase transition, noise, and external perturbation. Over the past decades, stochastic resonance has been extensively studied in systems of great diversity, from neuronal and brain functioning~\cite{Douglass1993Noise, Kitajo2003Behavioral} to semiconductor devices~\cite{Kittel1993Stochastic} and chemical reaction systems~\cite{Hohmann1996Stochastic}. In its essence, stochastic resonance is a result of the matching of two time scales: the period of the external perturbation and the transition rate. To observe the phenomenon of stochastic resonance a dynamical system needs to contain the following three basic ingredients~\cite{Gammaitoni1998}: (i) an energy barrier between metastable states; (ii) a weak coherent input, i.e., periodic external forcing, and (iii) a source of noise. For confined liquid crystals, it is not difficult to meet all the three conditions. In fact, a few previous works already tried to bring in the connection between stochastic resonance and liquid crystal systems~\cite{perc2008stochastic, Ho2012Fluctuation}. However, to our best knowledge, the relation between stochastic resonance and topological defects in a confined liquid crystal systems has not been investigated before. Moreover, because stochastic resonance is a prominent dynamical phenomenon that combines the effects of phase transition, noise, and external perturbation, it can be used as an ideal benchmark to test the dynamical properties of a liquid crystal system, which is what we do here.

In this paper we study the stochastic resonance phenomenon in the so-called planar bistable device in which nematic liquid crystals are confined in a shallow square well subject to planar boundary condition. Topological defects are unavoidable in this system because of the boundary constraint. First reported by Tsakonas et al.~\cite{tsakonas2007multistable}, this device features multiple defect-containing metastable states, including two ``diagonal'' solutions and four ``rotated'' solutions. In recent years this system has been extensively studied in the liquid crystal community. In the same paper~\cite{tsakonas2007multistable}, the authors were able to reproduced the observed multistable configurations using the Landau-de Gennes theory. Later, Luo et al.~\cite{Luo2012Multistability} studied the effect of boundary anchoring strength on the stability of the metastable states. Kusumaatmaja and Majumdar~\cite{kusumaatmaja2015free} computed the minimal energy paths connecting different metastable states over the Landau-de Gennes energy landscape. Using Oseen-Frank theory, Lewis et al.~\cite{Lewis2014Colloidal} obtained semi-analytic expressions for metastable states and realized them experimentally using filamentous virus particles. Within the molecular modeling framework, G\^{a}rlea and Mulder~\cite{G2014Defect} simulated long rod-like molecules in a square geometry using the Monte Carlo method. Slavinec et al.~\cite{Slavinec2015Impact} used a three-dimensional (3D) Lebwohl-Lasher model in a confined square well to show that the nematic structure is effectively two-dimensional (2D). Robinson et al.~\cite{Robinson2017} used several molecular models to study the fine structure of the metastable configuration.

While previous works have mainly focused on the static properties of the planar bistable device, here we aim to understand the dynamical properties of the system by studying the phase transition dynamics in the context of stochastic resonance. One interesting question is what kind of role topological defects play during the phase transition. Due to the mismatching of alignment direction of liquid crystal molecules near defects, maintaining a defect structure takes extra bending energy. Thus it is natural to ask, during the phase transition of the planar bistable device, will the pre-existing defect structures break down at the initial stage of a phase transition, or would remain intact until the final stage? Will new defect form during phase transition? Does the presence of defect promote or suppress the process of phase transition? As an initial attempt to address these questions, we try to link phase transition dynamics with topological defect by studying stochastic resonance in confined liquid crystals. In particular, we consider the planar Lebwohl-Lasher model~\cite{Lebwohl1972Nematic} with tangent boundary anchoring condition as a simple mathematical representation of the planar bistable device. Driven by a weak periodic external potential and intrinsic noise, the phase transition between the two ``diagonal'' solutions is simulated using an over-damped Langevin dynamics for different combinations of system size, boundary anchoring strength, amplitude of external potential, and noise level.

Our numerical results reveal that, driven by the oscillating external potential and noise, the system would take two different pathways during the phase transition process depending on the system size. As expected, stochastic resonance phenomenon can be observed for both transition pathways. However, the dependence of stochastic resonance to the boundary anchoring strength shows dramatic difference for systems taking different pathways. A remarkable feature we observed in relative small systems is the ``sticky-boundary'' effect, in which the phase transition gets stuck and stochastic resonance becomes de-activated when the boundary anchoring strength falls in a certain range. Our study demonstrates the intimate relation among geometry, noise, material properties, boundary effects and external fields during the phase transition dynamics in confined liquid crystals.

\section{Model and Numerical Methods}
We consider nematic liquid crystals confined in a 2D square subject to tangent anchoring condition at the boundary. We adopt the planar Lebwohl-Lasher model, which is the simplest model for nematic liquid crystals. The model consists of $N\times N$ particles whose positions are fixed on a regular lattice, plus another $4N$ particles forming the boundary (the four conner sites at the boundary are removed). The orientation of the $i$-th particle is confined in 2D space, and can be represented by the angle $\theta_i$ with respect to the $x$-axis. The Lebwohl-Lasher potential of the system is given by 
\begin{equation}
 H_{LL} = -\frac{1}{2}J\sum_{\{i,j\}_{nn}}P_2\big(\cos(\theta_i - \theta_j)\big),
 \label{eq_halmiltonina}
\end{equation}
where the summation is taken over nearest neighbors on the lattice. $J$ is the interaction strength. $P_2(x) = (3 x^2 - 1)/2$ is the second-order Legendre polynomial, and is introduced here to model the head-tail symmetry of the liquid crystal molecules, i.e., molecules pointing in the direction $\mathbf{n}$ and $-\mathbf{n}$ are indistinguishable.

The tangent anchoring condition is modeled by the boundary energy
$$
H_b = - \frac{1}{2} W \sum_{\{i\}_b} \cos^2(\theta_i - \theta_0),
$$ 
where the summation is taken for the boundary particles. $\theta_0$ is the corresponding tangent direction. $W$ is the anchoring strength.
To study stochastic resonance, a periodic external potential is applied to the system, which is given by
$$
H_e = -\frac{1}{2} A(t) \cos^2\left(\theta_i - \frac{\pi}{4}\right).
$$
The coefficient $A(t)$ is an oscillating function with period $T$ and amplitude $A_0$, 
\begin{equation}
A(t) = A_0\sin\left(\frac{2\pi t}{T}\right).
\label{eq_At}
\end{equation}
If $A(t) > 0$, the potential tends to align the particles along the main diagonal ($\pi/4$). On the other hand, when $A(t) < 0$, the potential tends to align the particles along the secondary diagonal ($3\pi/4$).
 
The total potential of the system is 
$$
H=H_{LL}+H_e + H_b.
$$
In our molecular dynamics simulation, the dynamics of the system is governed by the over-damped Langevin dynamics,
\begin{equation}
 \frac{d\theta_i}{dt} = \frac{\partial H}{\partial \theta_i} + \sqrt{2D}\xi_i(t). 
\end{equation}
Here $2D$ is the variance of Gaussian noise with zero mean and autocorrelation $\langle \xi_i(t)\xi_j(s) \rangle = \delta_{ij}\delta(t - s)$.

There are six parameters in the model, namely the system size $N$, the interaction strength $J$, the boundary anchoring strength $W$, the amplitude and period of the external field $A_0$ and $T$, and the noise level $D$. $J$ and $N$ are closely related to each other in the sense that decreasing $J$ gives similar results as increasing $N$ ($J$ determines the correlation length in the system). Thus we hold $J = 0.5$ fixed throughout our simulation. In addition, we fix $T=500$, which is much larger compared with the relaxation time of the system. The remaining four parameters are varied in our simulation to see how they affect the phase transition dynamics.


\section{Results}

\subsection{Metastable states}
\begin{figure*}
 \centering
 \includegraphics[width=.8\textwidth]{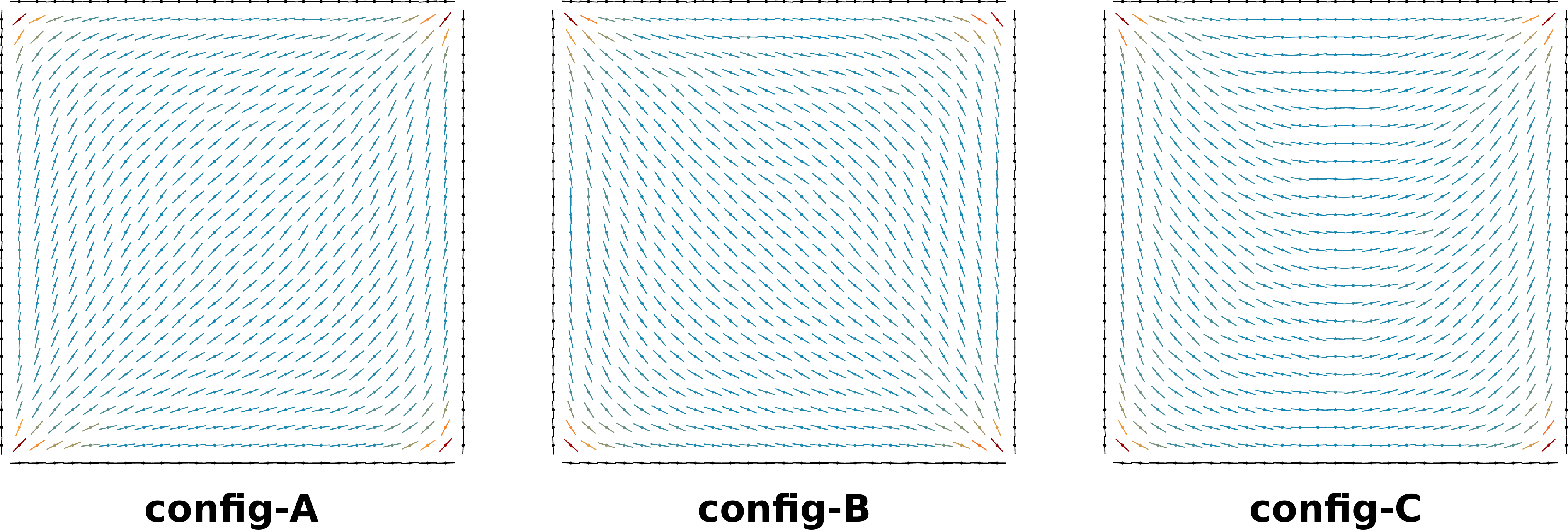}
 \caption{Metastable states obtained from simulation without external potential. The system size is $25 \times 25$ excluding particles at the boundary. Boundary particles are shown in black. The energy of each particle, normalized to $[-1, 0]$, is represented by color (blue means small energy and red means large energy).}
 \label{fig_states}
\end{figure*}

The metastable states observed in the experiments reported in~\cite{tsakonas2007multistable} can be recapitulated by our simulation. As shown in Fig.~\ref{fig_states}, when there is no external potential, we obtain the ``diagonal'' solutions (hereafter referred as config-A and config-B) as well as the ``rotate'' solution (config-C). The alignment direction in config-A roughly follows the main diagonal of the square, while in config-B follows the secondary diagonal. In config-A, there are two ``bending'' defects (particles pointing to the vertex) at the top-left and bottom-right corners and two ``splay'' defects (particles bending around the vertex) at the top-right and bottom-left corners. Rotating the config-A by $\pi$ one gets config-B. In config-C, the alignment direction bends like the alphabet ``C'' (with the opening facing upward). Rotating the config-C by $\pi/2$, $\pi$, and $3\pi/2$ gives rise to three other energetically equivalent ``C''-type configurations. In config-C, the two ``bending'' defects are connected by one edge of the square, which is also the case for the two ``splay'' defects. In our simulation, we obtain config-A, B, and C for a $25\times 25$ system (as shown in Fig.~\ref{fig_states}) and a $50\times 50$ system (results not shown). These results are consistent with~\cite{Robinson2017}.

\subsection{Transition paths}
In the following we consider the dynamical process of the phase transition between config-A and config-B, driven by external potential and intrinsic noise. The external potential $H_e(t)$ periodically switches between phase that favors config-A (when $A(t) > 0$) and phase that favors config-B (when $A(t)<0$). Such a periodic driving potential, aided by a proper amount of intrinsic noise, may significantly facilitate the phase transition between config-A and config-B through a mechanism called stochastic resonance. Note that the phase transitions between config-A and config-C or between config-B and config-C will not be considered in this paper. As reported in~\cite{Luo2012Multistability}, it is quite non-trivial to design an external potential that could drive such a diagonal-to-rotated transition and vice versa. 

\begin{figure*}
 \centering
 \includegraphics[width=.8\textwidth]{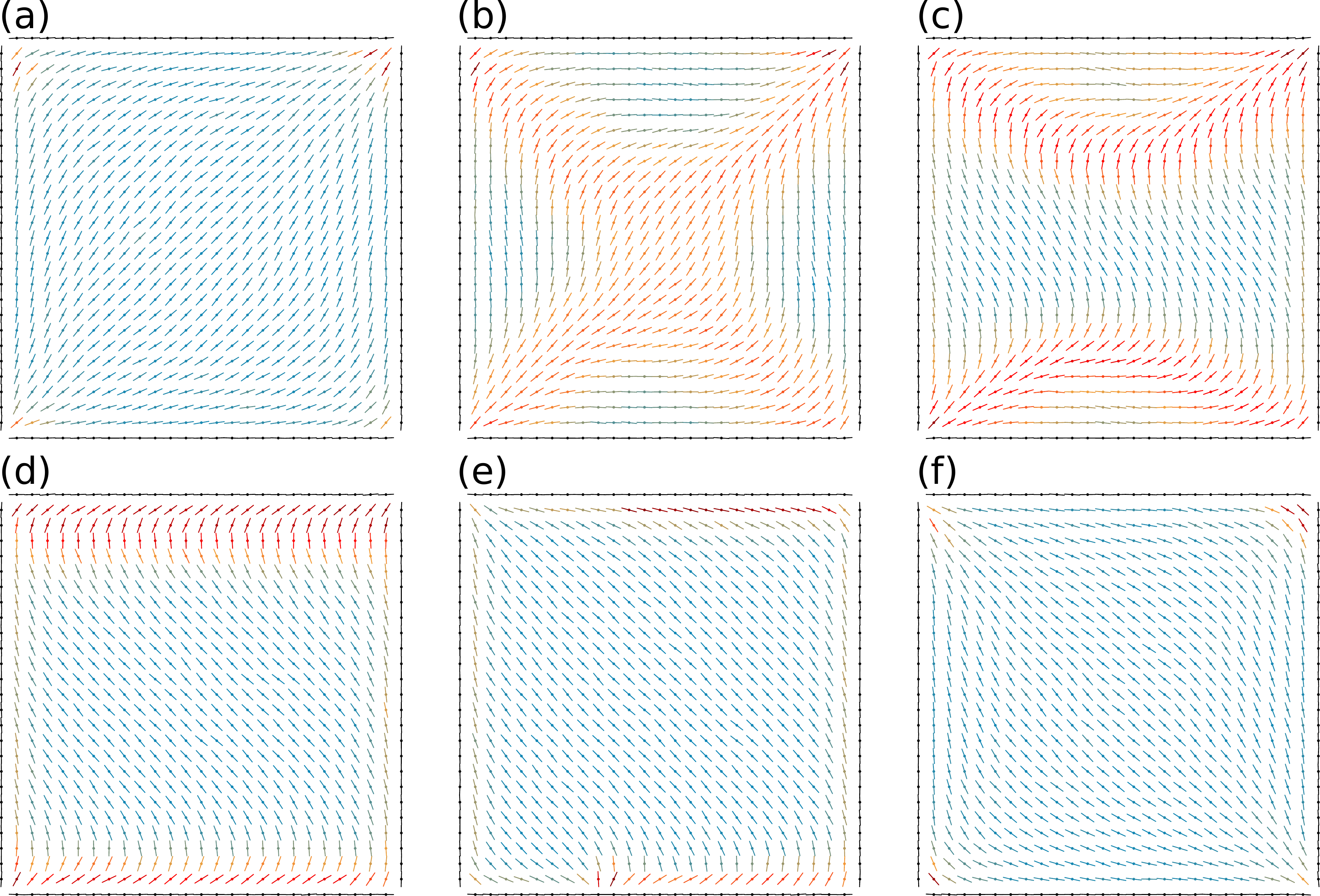}
 \caption{The ``S''-path from config-A to config-B driven by external potential. The system size is $25 \times 25$ excluding particles at the boundary. Boundary particles are colored in black, and internal particles are colored according to the normalized energy (blue for small energy and red for large energy). Parameters used are: $A_0=0.5$, $W=20$, $D=0.01$.}
 \label{fig_path1}
\end{figure*}

Before quantitatively measuring the phenomenon of stochastic resonance in our system, it is useful to investigate the phase transition dynamics, i.e., which pathway the transition process follows, and how it depends on the model parameters. In our simulation, we observe two distinct transition paths from config-A to config-B (the transition paths from config-B to config-A are the same under rotation symmetry). As shown in Fig.~\ref{fig_path1}, for the $25\times 25$ system, the first transition path typically starts with reorienting particles adjacent to the four edges. Meanwhile, particles next to the two square diagonals keep their original directions [see Fig.~\ref{fig_path1}(b)]. Under the external potential that favors config-B, these particles are in a state with relatively high energy. As the external potential strengthens, particles in the center of the square start to yield to it [Fig.~\ref{fig_path1}(c)], forming a continuous nematic region that extends to the square edges [Fig.~\ref{fig_path1}(d)]. Due to the boundary constraint, particles near the upper and lower edges of the square feel a high bending and external energy, which makes them want to flip over. The flipping-over process goes like a domino chain reaction along one edge, for which the triggering event may occur in one of the two vertices, or somewhere in the middle of the edge [Fig.~\ref{fig_path1}(e)]. After this, the system reaches config-B [Fig.~\ref{fig_path1}(f)]. Because this type of transition pathway needs to go through a characteristic transit state in which the alignment direction looks like the alphabet ``S'' [Fig.~\ref{fig_path1}(d)], we henceforth refer it as the ``S''-path. 

\begin{figure*}
 \centering
 \includegraphics[width=.8\textwidth]{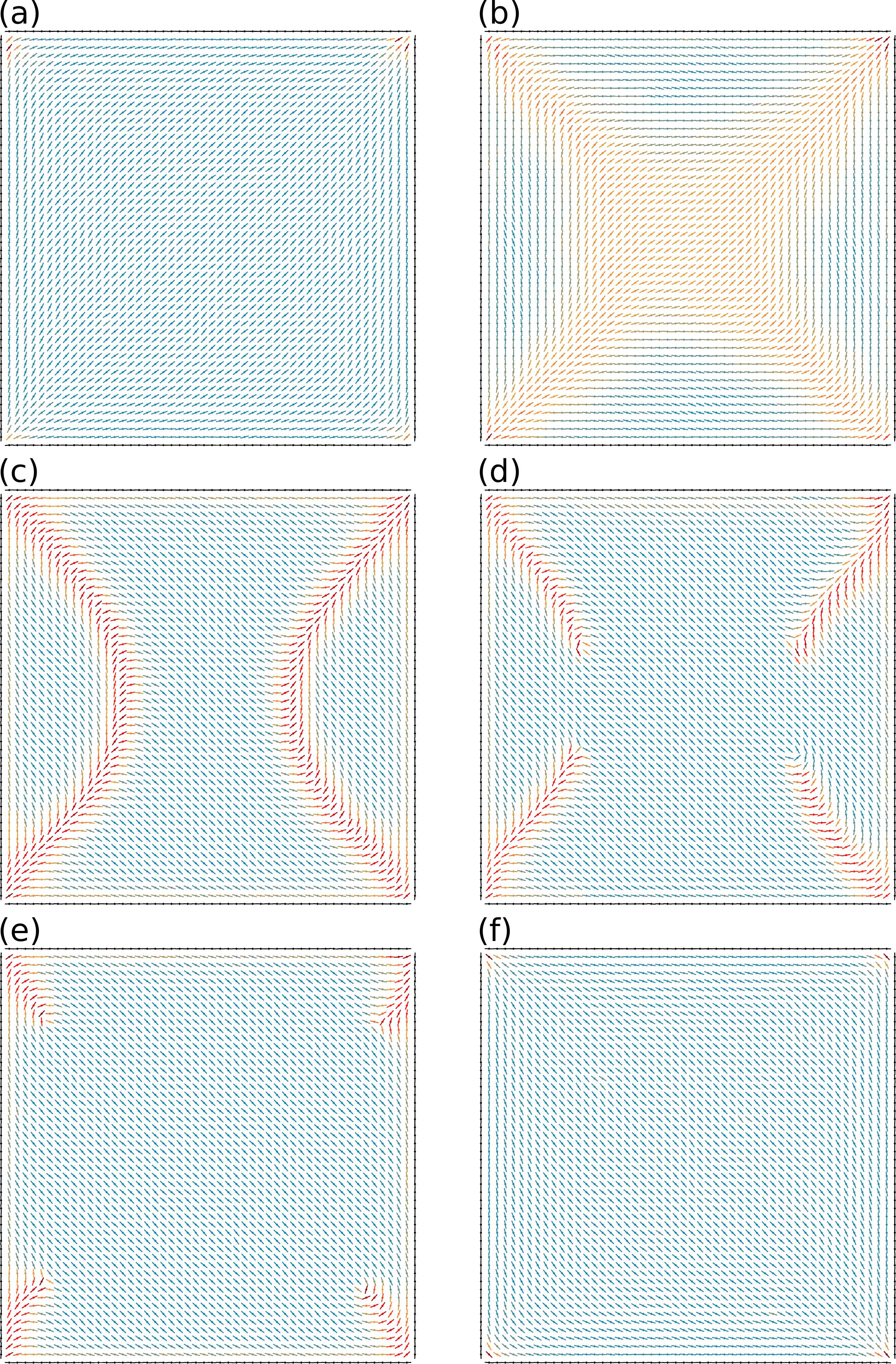}
 \caption{The ``X''-path from config-A to config-B driven by external potential. The system size is $50 \times 50$ excluding particles at the boundary. Boundary particles are colored in black, and internal particles are colored according to the normalized energy (blue for small energy and red for large energy). Parameters used are: $A_0=0.5$, $W=20$, $D=0.01$.}
 \label{fig_path2}
\end{figure*}

The other type of transition path is shown in Fig.~\ref{fig_path2} for the $50\times 50$ system. At the beginning it is similar to the ``S''-path till the two distorted lines (shown in red) form in Fig.~\ref{fig_path2}(c). However, unlike the ``S''-path for which the two distorted lines retreat to the square edges, here the two distorted lines break in the middle, giving rise to four newly formed point defects (two +1-defects and two -1-defects) inside the square [see Fig.~\ref{fig_path2}(d)]. These four point defects will gradually move to the corners [Fig.~\ref{fig_path2}(e)], where they merge with the pre-existing defects at the vertices. After merging, the system reaches config-B [Fig.~\ref{fig_path2}(f)]. One characteristic transit state along this type of transition pathway, as shown in Fig.~\ref{fig_path2}(d), contains four segments of distorted lines running in crossing directions inside the square, and we henceforth refer it as the ``X''-path. 

\begin{figure*}
 \centering
 \includegraphics[width=\textwidth]{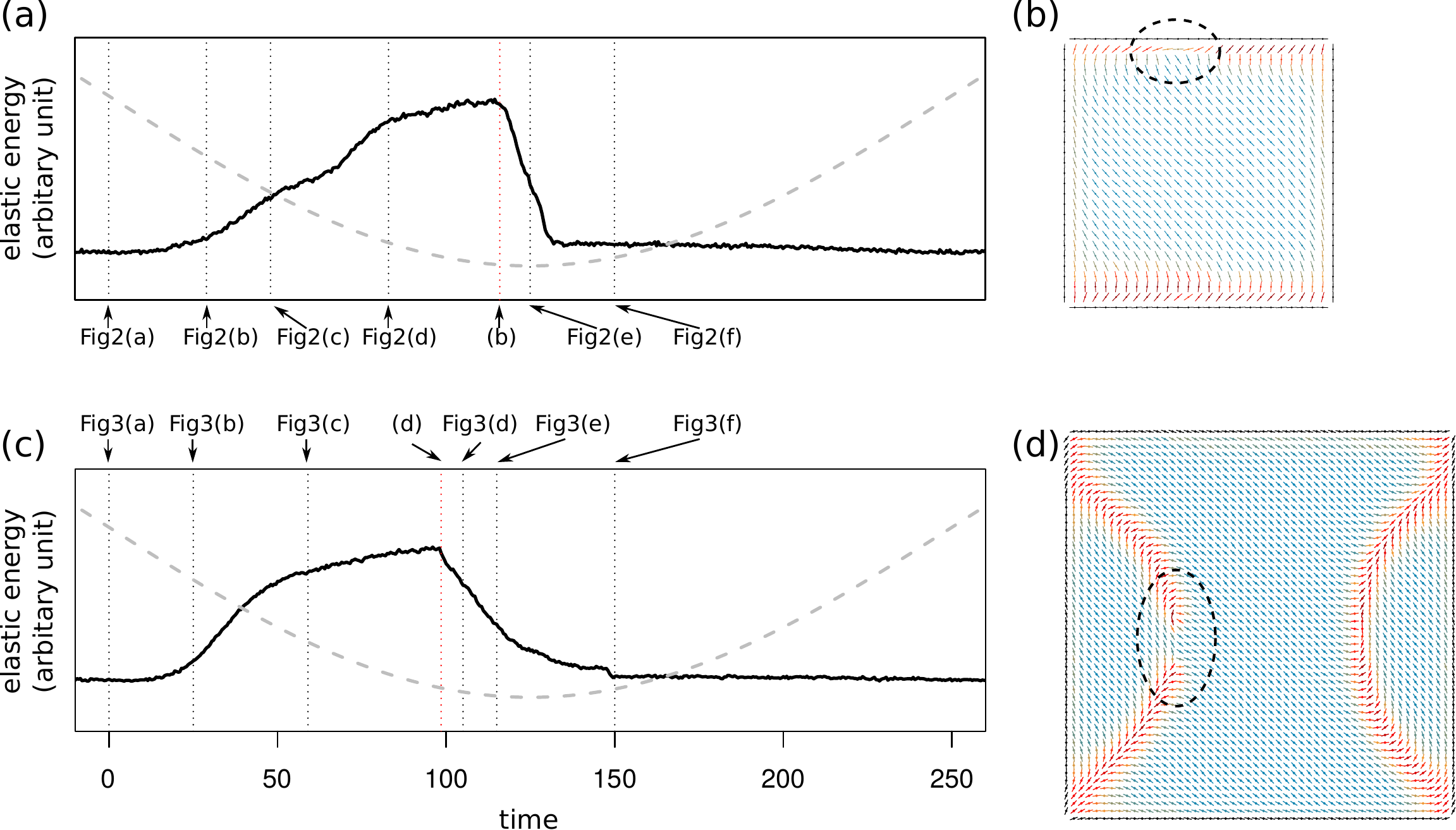}
 \caption{Time evolution of the elastic energy and critical points. (a) Elastic energy of the $25\times 25$ system (black curve). Black vertical dashed lines correspond to the states shown in Fig.~\ref{fig_path1}. Red vertical dashed line corresponds to the configuration with the highest elastic energy. Gray dashed curve indicates the strength of the external potential. (b) Critical point corresponding to the red vertical line in (a). (c) Elastic energy of the $50\times 50$ system (solid black curve). Black vertical dashed lines correspond to the states shown in Fig.~\ref{fig_path2}. Red vertical dashed line corresponds to the configuration with the highest elastic energy. Gray dashed curve indicates the strength of the external potential. (d) Critical point corresponding to the red vertical line in (c). Parameters used are: $A_0=0.5$, $W=20$, $D=0.01$.}
 \label{fig_energybarrier}
\end{figure*}

To highlight the difference between the two phase transition pathways, in Fig.~\ref{fig_energybarrier} we plot the time evolution of the elastic energy (Eq.~\eqref{eq_halmiltonina}) of the two systems. Starting from config-A at $t=0$, the external potential (represented by the gray-dashed curve in Fig.~\ref{fig_energybarrier}(a) and (c)) that favors config-B gradually increases its strength, and then decreases after reaching its maximum at $t=125$ (a quarter of the cycling period $T$). Driven by this external potential, the elastic energy in both cases (represented by the black curve in Fig.~\ref{fig_energybarrier}(a) and (c)) first increases and then decreases. We call the state when the elastic energy reaches maximum as the critical point (marked by the red-dashed vertical lines). These critical points can be considered as the rate-limiting step of the phase transition process. From Fig.~\ref{fig_energybarrier} (b) and (d), we can see that the critical points of the two systems are different. For the $25\times25$ system, the critical point corresponds to the collective reorientation of a number of molecules near the boundary (in the region marked by the black dashed circle), while for the $50\times 50$ system, the critical point corresponds to the breaking up of the distorted lines inside the rectangular. From Fig.~\ref{fig_energybarrier} we can see that the essential difference between the ``S''-path and ``X''-path is that, in the former the two distorted lines are pushed to the boundary while in the later the two distorted lines break up in the middle before they get pushed to the boundary. Whether or not the distorted lines would break up before being pushed to the boundary depends on a number of factors as discussed below.

\begin{figure*}
 \centering
 \includegraphics[width=.8\textwidth]{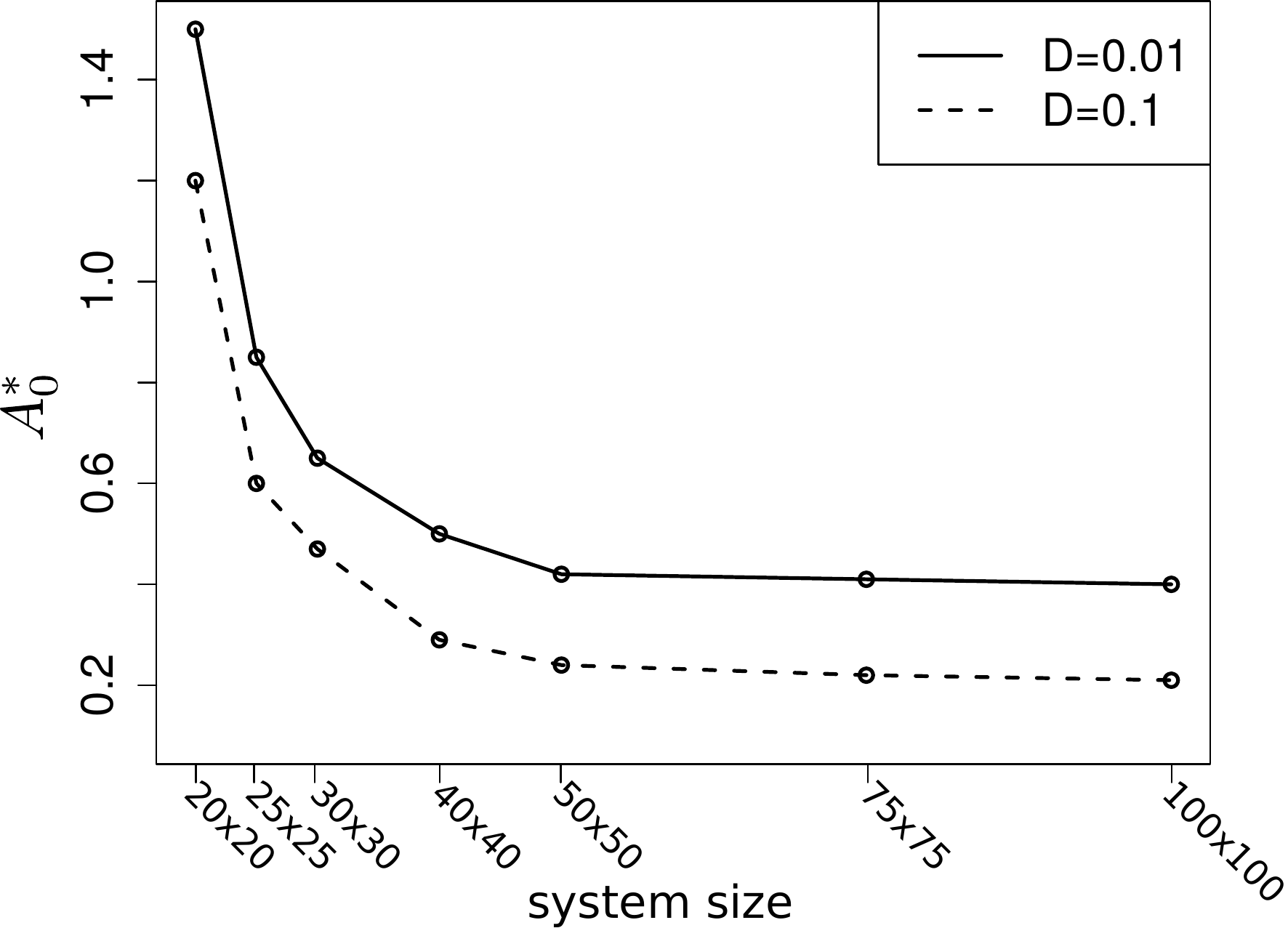}
 \caption{Threshold of external potential $A_0^*$ separating the ``S''-path and ``X''-path in systems with different sizes. Solid line corresponds to $D=0.01$ while dashed line corresponds to $D=0.1$. $W=20$.}
 \label{fig_A0star}
\end{figure*}

\begin{itemize}
 \item 
External potential $A_0$: As the driving force, the external potential needs to be large enough. For fixed system size and noise $D$, there exists a threshold $A_0^*$ for the amplitude of the external potential (Eq.~\eqref{eq_At}). If $A_0 \ge A_0^*$ the distorted lines will break up. Otherwise the distorted lines will not break up but get pushed to the boundary. 

\item 
Strength of noise $D$: For the same system size, $A_0^*$ tends to decrease as the strength of noise $D$ increases [see Fig.~\ref{fig_A0star}]. This is because noise can destabilize the distorted line structure and make it more prone to break. 

\item 
System size $N$: As we have seen in the previous examples that the $25\times 25$ and $50\times 50$ systems show different transition paths for the same $A_0$ and $D$. We speculate that this is a finite-size effect. Note that the breaking up of the two distorted lines would introduce four point defects in the bulk [Fig.~\ref{fig_path2}(d)], and the formation of such a structure in small system takes higher elastic energy due to the interaction between defects. To test this hypothesis, we record $A_0^*$ for systems with different size. If there is a finite-size effect, $A_0^*$ would approach to a constant as the systems size grows. We found that this is indeed the case [see Fig.~\ref{fig_A0star}]. Thus we conclude that ``S''-path is favored in systems with small size because of the finite-size effect.

\end{itemize}

Either path to follow, the transition from config-A to config-B contains two symmetry-breaking events. The first symmetry-breaking event corresponds to the separation of two distorted lines [Fig.~\ref{fig_path1}(b) $\rightarrow$ (c) or Fig.~\ref{fig_path2}(b) $\rightarrow$ (c)], which is the same for both the ``S''- and ``X''-path. However, the second symmetry-breaking events in the two types of transition pathways are different: in the ``S''-path it is the reorienting of particles at the boundary [Fig.~\ref{fig_energybarrier}(b)], while in the ``X''-path it is the break up of the distorted lines [Fig.~\ref{fig_energybarrier}(d)]. In both the ``S''- and ``X''-path, the second symmetry-breaking event is the rate-limiting step [see Fig.~\ref{fig_energybarrier}]. It is also worth noting that, if there is no noise or the noise is extremely small (on the order of $10^{-7}$), the first symmetry-breaking event is difficult to trigger. As a result, a ``X''-structure with perfect symmetry will form, which will break down in the center to give a structure like the one in Fig.~\ref{fig_path2}(d) for large enough $A_0$. Thus in the absence of noise, in both the $25\times25$ and $50\times 50$ systems the phase transition follows a special form of ``X''-path. We will come back to this point in the section of Discussions.

Later we shall see that, the dynamical properties in the $25\times 25$ and $50\times 50$ systems can be significantly different because of the different transition paths they follow.

\subsection{Spatial inhomogeneity}
In the previous subsection we have seen that the phase transition dynamics exhibits a clear spatial and temporal pattern. In this subsection we study how particles at different positions on the lattice respond differently to the external oscillating potential, which will provide some rational for us to quantify stochastic resonance for the whole system. 

First we define $s_i$ for each particle as
$$
s_i \equiv 2 \cos^2 \left(\theta_i - \frac{\pi}{4}\right) - 1.
$$
$s_i=1$ when the $i$-th particle is parallel to the main diagonal, and $s_i=-1$ when it is parallel to the secondary diagonal. Following~\cite{perc2008stochastic}, we then define $\gamma_i$ as,
\begin{eqnarray}
\gamma_i &=& \sqrt{\gamma_{i, \sin}^2 + \gamma_{i, \cos}^2}, \\
\gamma_{i, \sin} &=& \frac{2}{n T}\int_{0}^{n T} s_i(t) \sin\left(\frac{2\pi t}{T}\right) dt, \\
\gamma_{i, \cos} &=& \frac{2}{n T}\int_{0}^{n T} s_i(t) \cos\left(\frac{2\pi t}{T}\right) dt,
\end{eqnarray}
where $T$ is the period of the external potential, $n$ is the number of periods. $\gamma_i$ measures how significantly a single particle is in synchronization with the external oscillating potential.

\begin{figure*}
 \centering
 \includegraphics[width=.8\textwidth]{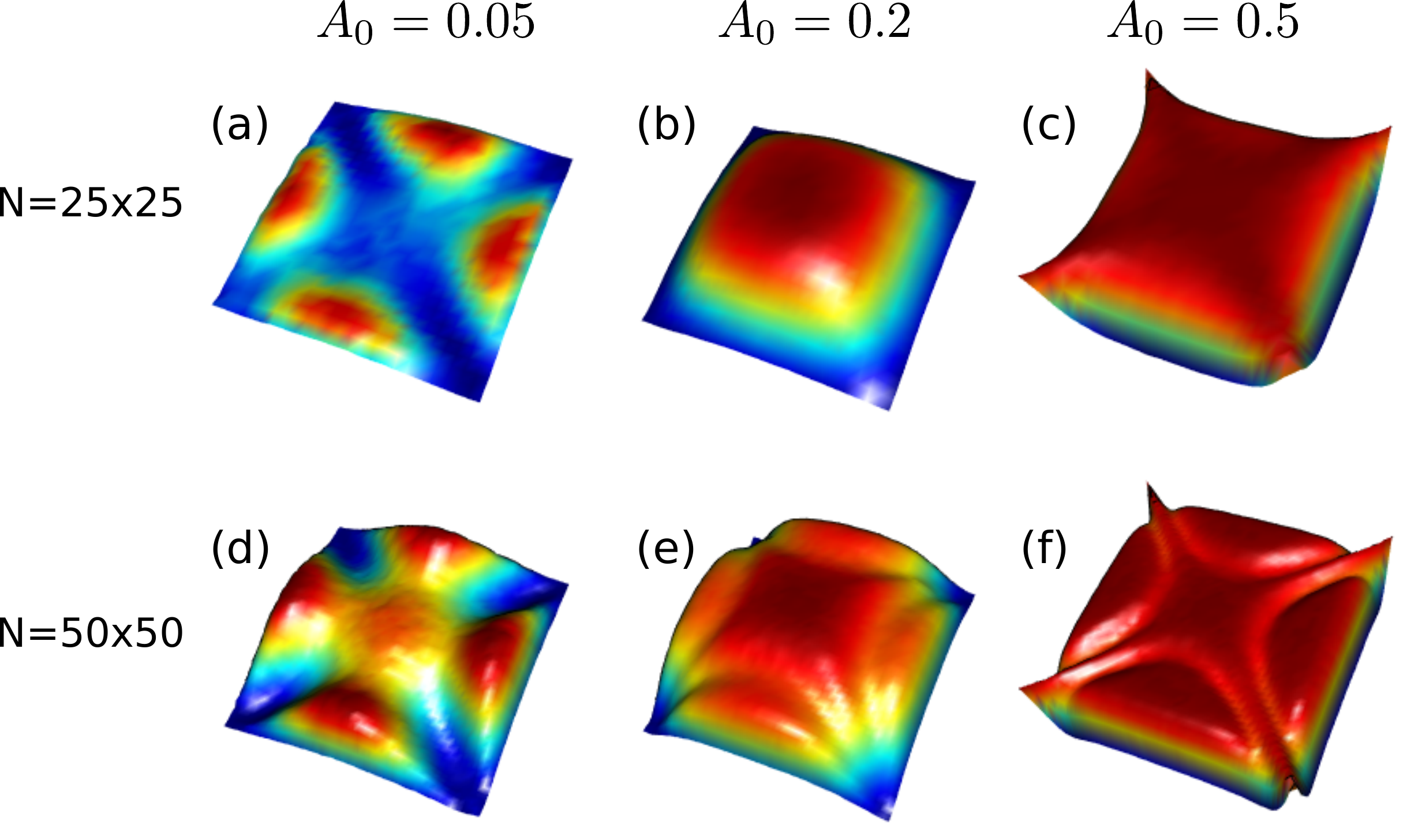}
 \caption{3D heatmap of the particle synchronization level in response to the oscillating potential. Both the altitude and the color represent the value of $\gamma_i$ at different particles in the square. (a-c) $25 \times 25$ system with $A_0 = 0.05, 0.2, 0.5$, respectively. (e-g) $50\times 50$ system with $A_0 = 0.05, 0.2, 0.5$, respectively. $W=20$, $D=0.01$, which are the same for both systems.}
 \label{fig_heatmap}
\end{figure*}

By changing the amplitude of the oscillating potential $A_0$ while keeping other parameters fixed, we compute $\gamma_i$ for each particle for the $25\times 25$ system and the $50\times 50$ system, respectively. The total simulation time is 10000, which contains $n=20$ oscillating periods. In Fig.~\ref{fig_heatmap} we plot $\gamma_i$ as 3D heatmap, in which hotter regions correspond to particles that are more active in response to the oscillating field, i.e., with larger $\gamma_i$. For the $25\times 25$ system [see Fig.~\ref{fig_heatmap}(a)-(c)], when $A_0$ is too small to trigger any symmetry-breaking event, only particles near the edges show slight response, while particles near the two square diagonals are largely steady. For moderate $A_0$ that can trigger the first symmetry-breaking event, yet still not big enough to trigger the second symmetry-breaking event to finish the whole phase transition, a large portion of particles in the center of the square show elevated synchronization level. For sufficiently large $A_0$ that grants the full config-A/B phase transition, particles at the corners become synchronized with the oscillating potential. These results are consistent with the spatial and temporal patterns of the corresponding phase transition pathway [see Fig.~\ref{fig_path1}]. For the $50\times 50$ system [see Fig.~\ref{fig_heatmap}(d)-(f)], there is a similar spatial inhomogeneity in the particles' response to the oscillating potential, although the heatmaps display a richer detailed structure. In particular, particles in the central region of the square become more active due to their relatively larger distance from the boundary and corner defects [see Fig.~\ref{fig_heatmap}(d)].

Together, these results suggest that the pre-existing topological defects are playing a role of stabilizing the macroscopic structure: the defects at the four corners of the square function like four nails fasten a tent, and the reorientation of particles at these defects marks the completion of phase transition for the whole system.

\subsection{Stochastic resonance}
Given that the hallmark of the config-A/B phase transition is the flipping of the particles at the four square corners, we thus define
\begin{equation}
 \Gamma = \gamma_1 + \gamma_2 + \gamma_3 + \gamma_4,
\end{equation}
to measure stochastic resonance of the system. Here index $i=1, 2, 3, 4$ correspond to the particles at the four square corners.

\begin{figure*}
 \centering
 \includegraphics[width=.8\textwidth]{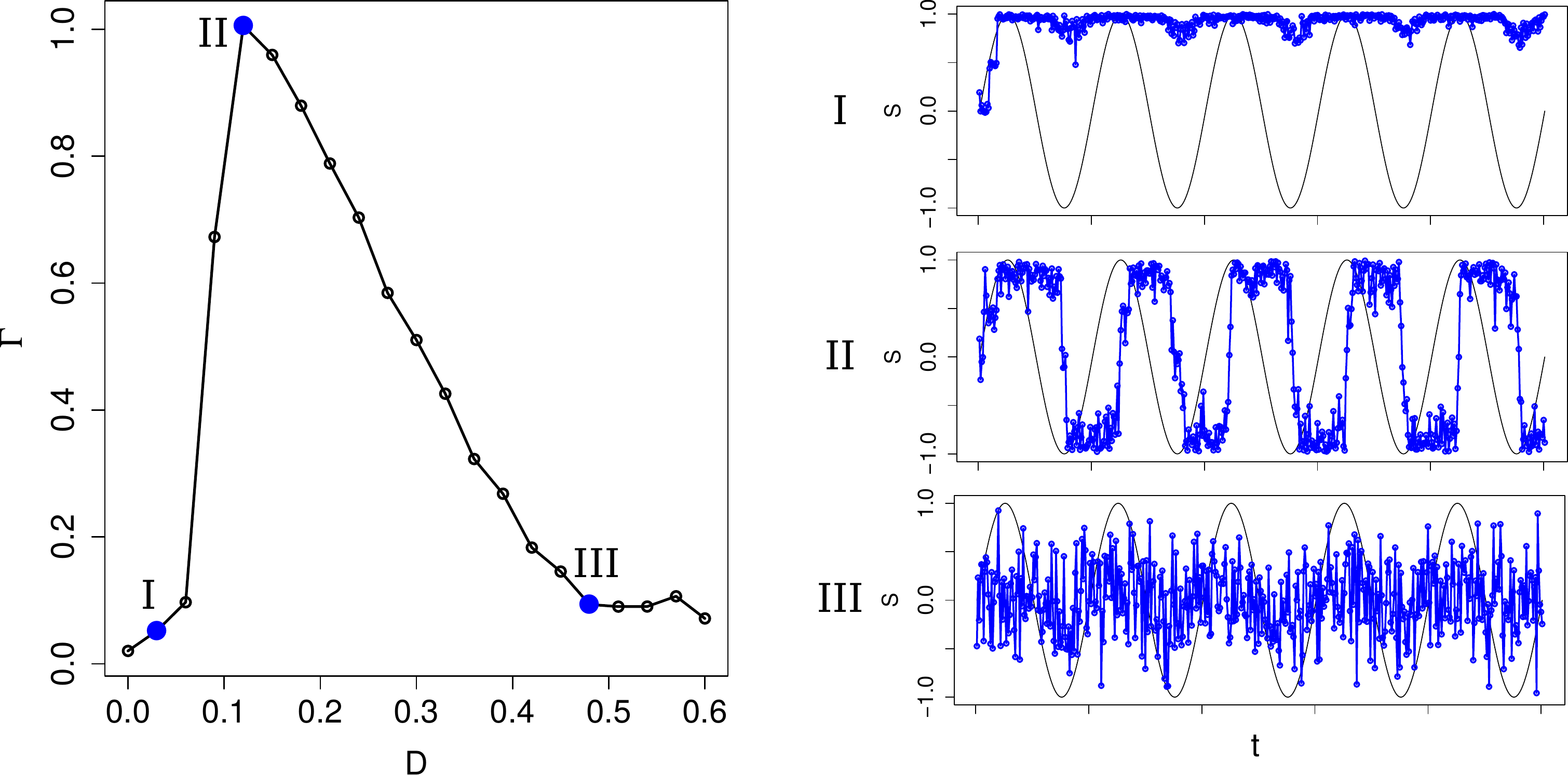}
 \caption{Left: $\Gamma$ vs $D$. Right: $S$ (defined as $s_1 + s_2 + s_3 + s_4$) vs time $t$. I, II, and III correspond to the three points marked on the left. Black line represents the oscillating external potential. System size: $25\times 25$, $W=20$, $A_0=0.35$.}
 \label{fig_QvsD}
\end{figure*}

To test whether stochastic resonance occurs in our system, we apply a relatively weak external potential to compute $\Gamma$ for different noise level $D$. By definition, we say the phenomenon of stochastic resonance is observed if there is a non-monotonic relationship between $\Gamma$ and $D$. As shown in Fig.~\ref{fig_QvsD}, stochastic resonance indeed exists in our system. In particular, when the noise is too small (I), there is no phase transition. When the noise is too large (III), the particle orientation becomes almost isotropic. Only when there is a right amount of noise so that the phase transition rate of the system matches with the oscillating rate of the external potential, the system shows full synchronization with the external potential (II), i.~e., resonance occurs. 

\begin{figure*}
 \centering
 \includegraphics[width=.8\textwidth]{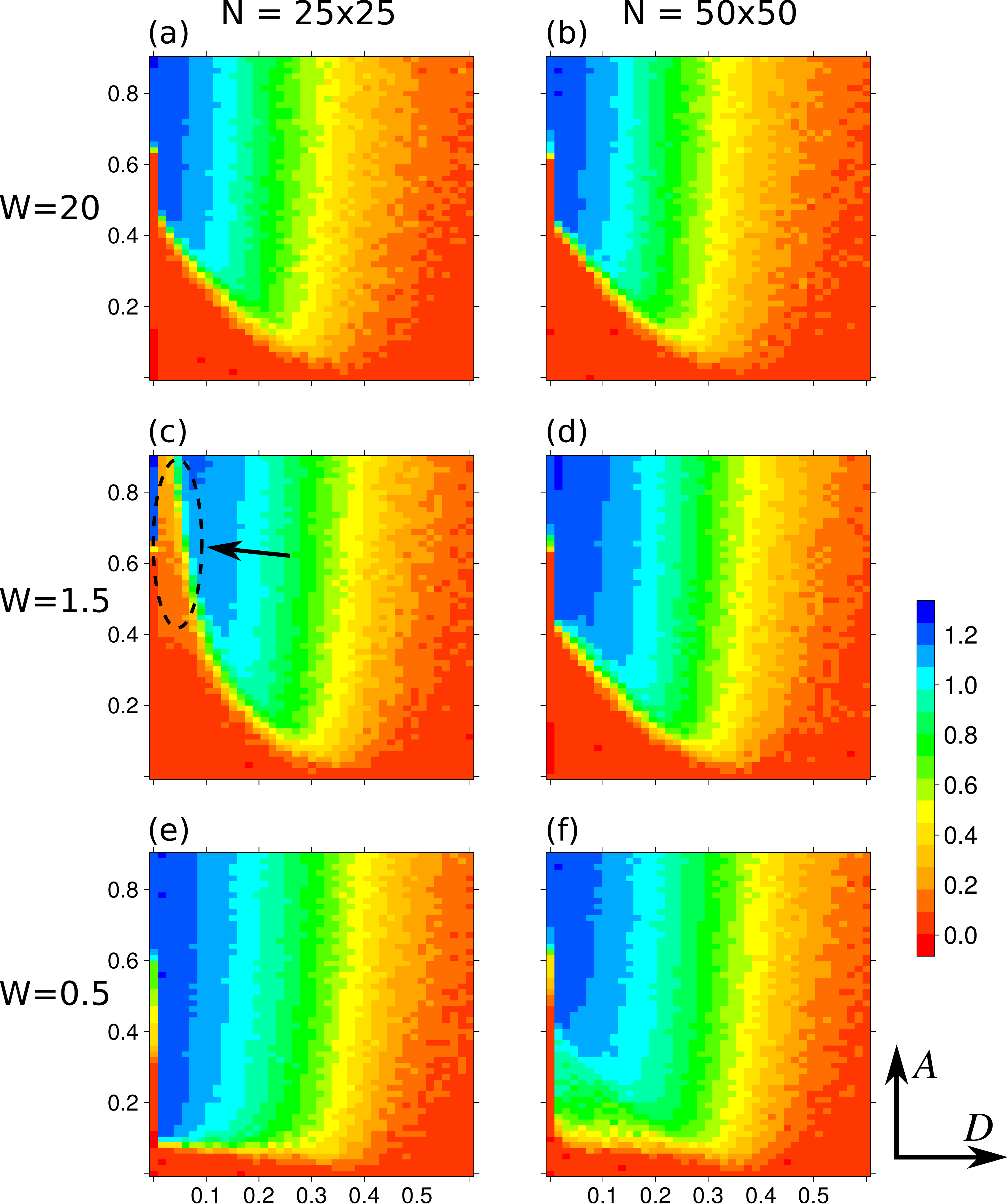}
 \caption{$\Gamma$ for different parameters. Each panel corresponds to $\Gamma$ obtained for different $D$ and $A_0$, with the value of $\Gamma$ shown in color. (a, c, e) correspond to the $25\times 25$ system and (b, d, f) correspond to the $50\times 50$ system. (a, b) $W=20$. (c, d) $W=1.5$. (e, f) $W=0.5$. For each set of parameters, the simulation time is $t=10000$, which contains $n=20$ oscillating periods.}
 \label{fig_SRmap}
\end{figure*}

Next we investigate how stochastic resonance depends on model parameters, including the amplitude of the external potential $A_0$, the boundary anchoring strength $W$, the noise level $D$, and the system size $N$. We compute $\Gamma$ for each set of parameters. These results are presented in Fig.~\ref{fig_SRmap}. For the $25\times 25$ system with a relatively large anchoring strength $W=20$ [see Fig.~\ref{fig_SRmap}(a)], stochastic resonance can be observed for $A_0$ ranging from 0.05 to 0.6. The noise level corresponding to the peak value of $\Gamma$ decreases as $A_0$ increases, which can be explained by the fact that the energy barrier between config-A and config-B decreases as the external potential increases. For $A_0 > 0.6$, stochastic resonance disappears and $\Gamma$ decreases monotonically as $D$ increases, suggesting that the energy barrier is gone. For $D>0.5$, the liquid crystals become almost isotropic due to large noise. If we change the system size from $25\times 25$ to $50\times 50$, the heatmap is almost the same for strong boundary anchoring [Fig.~\ref{fig_SRmap}(b)], even though the phase transition map actually follow a different pathway [see Fig.~\ref{fig_path1} and~\ref{fig_path2}]. However, if we reduce the boundary anchoring strength $W$ to 1.5, the heatmaps obtained from the $25\times 25$ system and the $50\times 50$ system show significant difference for $D$ around 0.05 [indicated by the region pointed by the arrow in Fig.~\ref{fig_SRmap}(c)]. Apparently the phase transition becomes much harder as a result of reduced boundary anchoring in the $25\times 25$ system, but this phenomenon does not occur in the $50\times 50$ system. The reason is that, for this boundary anchoring strength, the $25\times 25$ system gets stuck in the ``S''-configuration [see Fig.~\ref{fig_path1}(d)]. It could not proceed to finish the phase transition because the boundary anchoring strength is not strong enough to provide enough bending energy, and at the same time not weak enough to release the nearby particles so that they can reorient to yield to the external potential. We call this intriguing dynamical property as the ``sticky-boundary'' effect, and it is a cooperative result of the boundary anchoring, intrinsic noise, and external potential. Interestingly, it only happens for phase transition that follows the ``S''-path. There is no ``sticky-boundary'' effect in the $50\times 50$ system [see Fig.~\ref{fig_SRmap}(d)] because the phase transition follows the ``X''-path, in which the second symmetry-breaking event occurs in the bulk rather than at the boundary [see Fig.~\ref{fig_energybarrier}]. For small anchoring strength ($W=0.5$), the ``sticky-boundary'' effect disappears in the $25\times 25$ system, and the phase transition becomes much easier for a much wider range of $A_0$ and $D$ [Fig.~\ref{fig_SRmap}(e)]. For the $50\times 50$ system, the trend that phase transition becomes easier as $W$ decreases always holds [Fig.~\ref{fig_SRmap}(f)]. Note that in the limit of $W=0$, for both systems there will be no energy barrier separating config-A and config-B, as long as there exists a non-zero external potential that favors one configuration. 

\section{Discussions}
In this paper, we study the phase transition dynamics driven by external forcing and noise in a confined liquid crystal system. It is a simple multistable system that contains topological defects. As expected, stochastic resonance can be observed in this system. We quantify stochastic resonance using different combinations of system size, boundary anchoring strength, and external potential amplitude. Interestingly, the quantitative behavior of stochastic resonance can be dramatically different depending on the phase transition pathway the system follows. A few key questions about the intimate interplay among topological defect, boundary constraint, external forcing, and noise during the phase transition dynamics are discussed below.

\subsection*{Topological defect and phase transition}
Now we address the question raised earlier in the Introduction --- does the presence of defect promote or suppress phase transition? In fact, both effects have been observed in our numerical simulation. On one hand, as shown in Fig.~\ref{fig_heatmap}, the pre-existing ``bending'' and ``splay'' defects appear to be stabilizing the macroscopic structure, and particles near these defects do not switch their orientations until the last stage of the phase transition. On the other hand, when the phase transition follows the ``X''-path, new defect points emerged inside the square appear to promote the phase transition process: particles around these newly formed point defects reorientate and ``push'' the defect points to the corners. Guided by these observations, we believe that defects are merely local structures in a metastable or transit states. Whether or not the presence of defect promotes or suppresses phase transitions in a confined liquid crystal system appears to be dependent on the transition pathway. The transition pathway, however, depends on the whole energy landscape and thus resides on a more global level than defects themselves.

\subsection*{Sticky-boundary effect}
In our system, the energy barrier between config-A and config-B is a result of the boundary constraint (in the absence of the boundary constraint and external potential, the alignment direction of the nematic liquid crystals can rotate continuously from config-A to config-B without changing the Lebwohl-Larsher energy of the system). Our initial guess is that the phase transition process would become easier as the boundary anchoring strength decreases. In fact, for the same planar bistable device, using the Landau-de Gennes energy and doubly-nudged elastic band method~\cite{Trygubenko2004A}, Kusumaatmaja and Majumdar~\cite{kusumaatmaja2015free} computed the minimal energy paths between config-A and config-C (thus different from the config-A/B phase transition considered here, and also note that our system has an external potential while theirs do not). They found that the energy barrier along the minimal energy path decreases monotonically as boundary anchoring strength $W$ decreases. However, to our surprise, in our case we observe that under certain conditions the phase transition for $W=1.5$ is even harder than that of $W=20$ [see Fig.~\ref{fig_SRmap}(c)]. This phenomenon is path dependent: it only occurs for the ``S''-path in which the distorted lines are pushed to the boundary. The reason that the phase transition is suppressed is that the relative weak boundary anchoring ``absorbs'' the bending energy of particles near the boundary. We call this counter-intuitive phenomenon as the ``sticky-boundary'' effect.

\begin{figure}
 \centering
 \includegraphics[width=.5\textwidth]{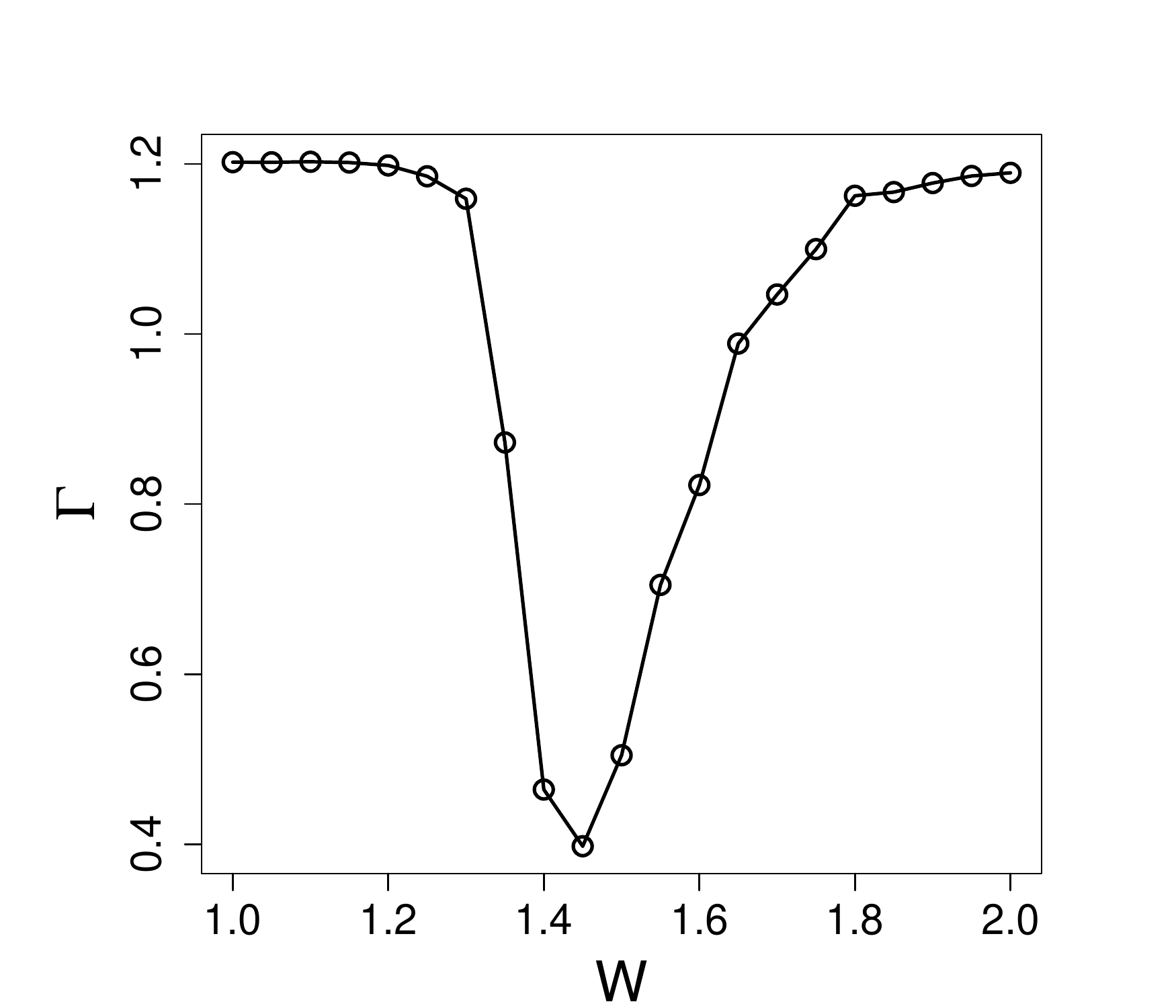}
 \caption{$\Gamma$ vs $W$. System size is $25\times 25$. $A=0.7$. $D=0.05$. The simulation time is $t=10000$, which contains $n=20$ oscillating periods.}
 \label{fig_QvsW}
\end{figure}

Because of the ``sticky-boundary'' effect, the dynamical response of the system $\Gamma$ may exhibit a non-monotonic relationship with the boundary anchoring strength $W$ [see Fig.~\ref{fig_QvsW}]. In another work, Luo et al.~\cite{Luo2012Multistability} reported that selectively applying variable anchoring strength on certain parts of the boundary may mediate phase transition in the planar bistable device. Together, these novel dynamical properties in confined liquid crystals under various boundary anchoring conditions may be useful in designing programable devices in practice.

\subsection*{The effect of noise}
In both the $25\times 25$ and $50\times 50$ systems, if there is no noise or the noise is extremely small (on the order of $10^{-7}$), the first symmetry-breaking event is much harder to occur compared with the case with noise. This is because when the external potential is relatively weak, the particles on different sides of the square diagonals are bending towards opposite directions [see Fig.~\ref{fig_path1}(b)]. If noise exists, the ``X''-structure running across the square diagonals will not be in perfect symmetry (for example, in Fig.~\ref{fig_path1}(b) the ``X'' is slightly elongated in the vertical direction). If there is no noise, however, the ``X''-structure will be in perfect symmetry. As the external potential gradually increases, the alignment of particles on the ``X''-structure bends more severely. If the external potential exceeds a certain threshold value, the ``X''-configuration would break down in the center, generating four point defects, and the two symmetry-breaking events in the ``X''-path appears to occur simultaneously. After this point the phase transition pathway looks like the ``X''-path as in Fig.~\ref{fig_path2}(d)-(f). Note that the distorted lines do not retreat to the boundary as in the ``S''-path, which explains why there is no ``sticky-boundary'' effect for $D=0$ in Fig.~\ref{fig_SRmap}(c).

To summarize, in our system the effect of noise in phase transition dynamics can be quantitative and qualitative. Quantitatively, increasing the strength of noise can speed up the phase transition process. Qualitatively, with and without noise the phase transition may take completely different pathways.

The phenomenon that the phase transition takes different pathways under the influence of noise might be also true for other spatial-temporal dynamics on a complex energy landscape. Conventional methods of computing the minimal energy path, such as the string method~\cite{weinan2002string} or the doubly-nudged elastic band method~\cite{Trygubenko2004A}, that do not take noise into account may not be able to reveal the actual transition path for systems like ours in the presence of noise. In future it would be interest to investigate the noise-dependent phase transition dynamics using our simple system as a toy model.

\subsection*{Choice of liquid crystal model}
In this paper we study the lattice-based Lebwohl-Lasher model, which is the simplest model of liquid crystals. For the planar bistable device, in~\cite{Robinson2017} two other molecular models are used, which are the Hard-Gaussian-Overlap model and the Gay-Berne model. Both of these two models are off-lattice, which means that the molecular position is another degree-of-freedom in addition to the molecular alignment direction. Another modeling framework is the continuous model, in which the system is described by the macroscopic order parameter, such as the $Q$-tensor. For example, in~\cite{Ho2012Fluctuation} the Oseen-Frank model is used and a Langevin dynamics with a spatial-temporal correlated Gaussian noise is derived. The benefit of using a continuous model is that a partial differential equation can be derived, and there are many efficient solvers could be used. However, the physical justification for the noise terms is less clear compared with the molecular model. In future, it would be interest to compare the properties of stochastic resonance in different models.

\subsection*{Future work and connection with experiments}
The model considered here may be oversimplified in terms of quantitatively predicting the experimental result. However, we believe some qualitative features, such as the different pathways dependent on system size and noise, and the ``stick-boundary'' effect can be also observed in other models and also in experiments. It would be interesting to investigate more complex models such as the Landau-de Gennes model. Current experiment about the bistable planar device focus on the static properties of the device. It would be interesting to directly observe the dynamical process such as the phase transition and see the validation and limitation of the model.

\begin{acknowledgments}
YH and LH are supported by National Science Foundation of China (grant 11671415) and Tsinghua University Initiative Scientific Research Program (grant 20151080424). We thank the anonymous reviewers for their helpful suggestions.
\end{acknowledgments}

\bibliography{ref}
\end{document}